# Wireless energy transfer between anisotropic metamaterials shells

Ana Díaz-Rubio, Jorge Carbonell, and José Sánchez-Dehesa*

*Wave Phenomena Group (GFO), Departamento de Ingeniería Electrónica (7F), Universitat Politècnica de Valencia, Camino de Vera, s/n, 46022 Valencia, Spain*

**Abstract**

The behavior of strongly coupled Radial Photonic Crystals shells is investigated as a potential alternative to transfer electromagnetic energy wirelessly. These sub-wavelength resonant microstructures, which are based on anisotropic metamaterials, can produce efficient coupling phenomena due to their high quality factor. A configuration of selected constitutive parameters (permittivity and permeability) is analyzed in terms of its resonant characteristics. The coupling to loss ratio between two coupled resonators is calculated as a function of distance, the maximum (in excess of 300) is obtained when the shells are separated by three times their radius. Under practical conditions an 83% of maximum power transfer has been also estimated.

* author to whom correspondence should be addressed: jsdehesa@upv.es

# 1. Introduction

Wireless energy transfer has received in the last few years a renewed interest partly linked to the tremendous development of mobile and wireless devices. Transmitting energy without the need of a connecting physical infrastructure was already a matter of study at the times of Nicola Tesla [1], or even Heinrich Hertz. Nevertheless, the advent of the electrical wired grid has hidden that interest for a long time. More recently, the evident advantages of this type of technology have made it especially attractive in certain environments or circumstances. For example, mobile devices of all sorts (telephones, robots, sensors, cars…) or devices working in harsh environments with access limitations (satellites or emergency systems) would sensibly benefit from the possibility of using wireless power to recharge their batteries or allow their un-interrupted operation without severely compromising mobility. Different types of technological schemes have been proposed to perform a wireless energy transfer without the need of a physical connection [2-4]. In this sense, radiative and non-radiative energy transfers have been investigated for metamaterial structures [5], and resonant coupling between relatively close resonant elements has been widely studied from different points of view [6,7]. This phenomenon takes advantage of a strong electromagnetic coupling between resonant elements that fundamentally have a sub-wavelength size. At the same time, resonators with high quality factors ($Q$) are required so that absorptive and radiative losses may not preclude an effective transfer of energy from the source element to a load element. In reaching high efficiencies, examples of coil-based systems have been widely investigated, but they have the drawback of being bulky, especially when working at low frequencies [8,9].

In this letter we propose the use of Radial Photonic Crystals (RPCs) shells as resonant elements enabling electromagnetic energy transfer wirelessly between them. The proposed shells consist of a layered anisotropic metamaterial whose constitutive parameters (namely electric permittivity $\varepsilon$ and magnetic permeability $\mu$) are radially dependent. They have a cylindrical shape with a central void cavity. Among other purposes, valid also for their acoustic counterparts [10,11], RPC shells have been employed as high $Q$-factor resonant elements, position and frequency sensors or structures for beam-shaping [12,13]. These functionalities are a consequence of the anisotropy of the constitutive parameters. RPCs shells can be considered 'very ordered' systems and this fact is also linked to a high degree of flexibility in implementing their resonant characteristics, both in terms of resonant mode selection as well as in the operation frequency range (due to scalability). Other advantages are pointed out in the following.

We start by introducing a design of a metamaterial shell with the target of having a high-$Q$ resonance at low (microwave) frequencies. The design is based on a 4-layer cylindrical microstructure of sub-wavelength size at the operation frequency. A particular operation mode is selected so that the analysis of a system formed by two coupled shells allows the evaluation of the energy transfer between them. A strong coupling regime is established when one of the two identical shells acts as a source and the other acts as the load of the system [14]. The shells are analyzed in terms of their quality $Q$ and coupling $\kappa$ factors, which are compared with other possible solution implemented with homogeneous materials. This comparison is performed through a figure of merit (FOM) relating loss and coupling. Finally, the performance of the energy transfer is evaluated by means of an efficiency figure and discussed in view of possible implementation of this proposal.

## 2. Theoretical background

The designed anisotropic metamaterial shell consists of alternating layers of type $a$ and of type $b$ whose constitutive parameters are as follows:

$$\mu_{ra}(r) = \frac{40d}{r}; \qquad \mu_{rb}(r) = \frac{60d}{r}, \qquad (1)$$

$$\mu_{\theta a}(r) = \frac{35d}{r}; \qquad \mu_{\theta b}(r) = \frac{20d}{r} \qquad (2)$$

$$\varepsilon_{za}(r) = \frac{40d}{r}; \qquad \varepsilon_{zb}(r) = \frac{60d}{r} \qquad (3)$$

These constitutive parameters have been chosen in order to make the radial impedance $Z_r(r) = \sqrt{\frac{\varepsilon_z(r)}{\mu_r(r)}}$ equal to the vacuum impedance. This design facilitates the energy transfer between the shells.

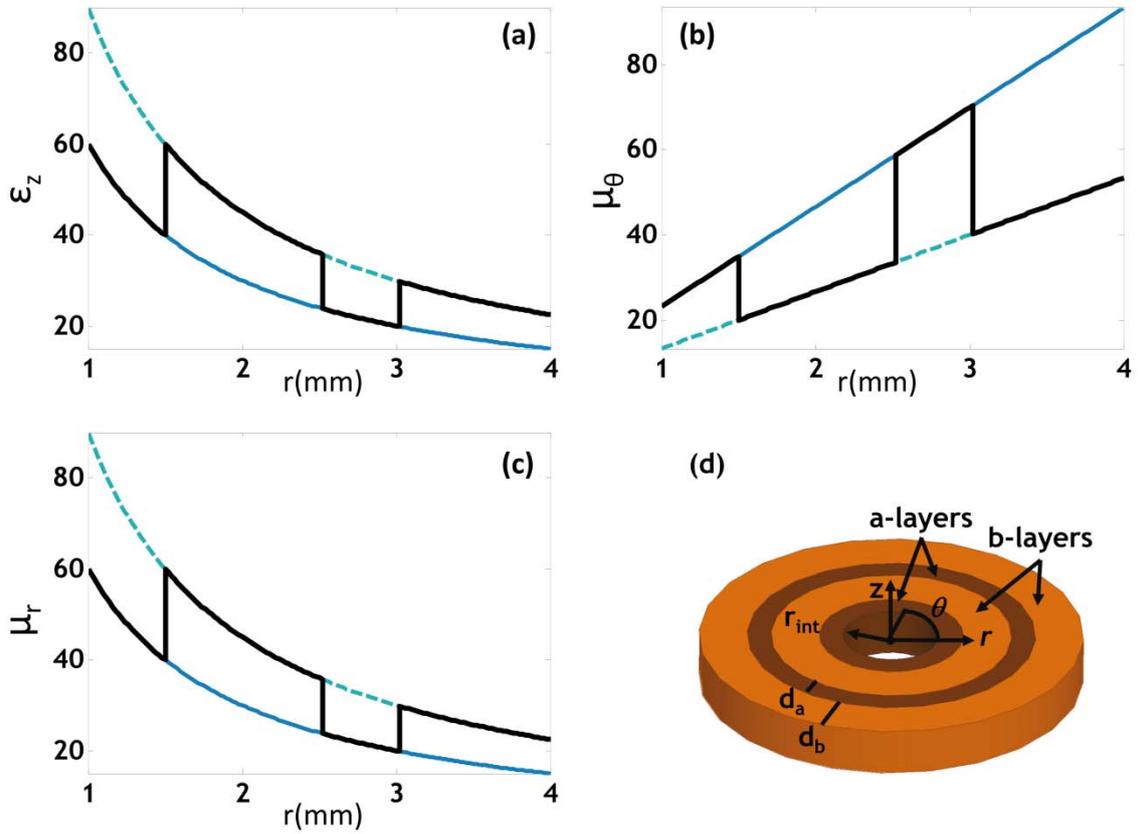

Fig. 1. Constitutive parameters as a function of the radial distance from the center of the RPC (a) radial permeability profile $\mu_r$, (b) angular permeability profile $\mu_\theta$, and (c) electric permittivity $\varepsilon_z$. Thick line represents the actual parameter profile alternating and bounded between the continuous (thin lines) profiles of each layer type (*a*-type or *b*-type). In (d), schematic view of a 4-layer RPC, including a void central cavity and with characteristic dimensions: $r_{int}$ = 10 mm, $d_a$ = 5 mm, $d_b$ = 10 mm. Radial periodicity is $d = d_a + d_b = 15$ mm.

Figure 1 displays the functions above for a 4-layer resonator. The stepped radial dependence of all parameters produces a resonant structure with multiple modes of different symmetries. The moderately-high permittivity and permeability values shown in Fig. 1 (< 60) can be obtained in practice by using metamaterial inspired elements as in previous works [12,13] or using combinations of natural materials with these permittivity or permeability values (in the MHz range). Alternative approaches can even be based on the use of discrete elements as in [15], when working at sufficiently low frequencies and with moderate physical sizes of a few centimeters.

Note that the parameter values described in Fig. 1 are not extreme as the ones proposed in [6] for dielectric structures, where $\varepsilon_r$ = 147.7. Moreover, also note that using a complex profile of permittivity and permeability allows reducing the resonant frequencies of the shell at which both types of energy, electric and magnetic, are collected. Remember that the total energy density in a single shell is:

$$W_{tot} = W_E + W_M = \frac{1}{2}\varepsilon E^2 + \frac{1}{2\mu}H^2$$

The goal with the permittivity and permeability profiles is generating a small electrical size resonator working at a frequency around 179.33 MHz. As an example, the schematic shell depicted in Fig. 1(d), has a radial periodicity parameter $d$ = 15 mm and an external radius $r_{ext}$ = 40 mm, and works under a quadrupolar resonance (see Fig. 2(a)). It can be easily demonstrated that the shell has a sub-wavelength size at the operation frequency; i.e., $\lambda/r_{ext} \approx 42$. The sub-wavelength size is crucial to operate in a strong coupling regime since evanescent components of the resonant fields can also be used to increase the energy transfer. Using a long-wavelength regime makes that moderate separations between two shells (< $10r_{ext}$) can be smaller than a free space wavelength and hence energy in the rapidly decaying evanescent components can be effectively trapped by the load device.

The single metamaterial shell has been analyzed employing Comsol, [16], a software package that performs 2D numerical simulations and identifies its resonant frequencies. It can be shown that resonance frequencies and mode symmetries of the finite structure can be derived from the dispersion diagram of the infinitely periodic structure with same definition equations ((1) to (3)), [10].

## 3. Results and Discussion

Resonant modes with different symmetry order are found at $f_{q=1}$ = 143.91 MHz, $f_{q=2}$ = 179.33 MHz and $f_{q=3}$ = 223.87 MHz; value of $q$ stands here for the dipolar, quadrupolar or sextupolar order respectively.

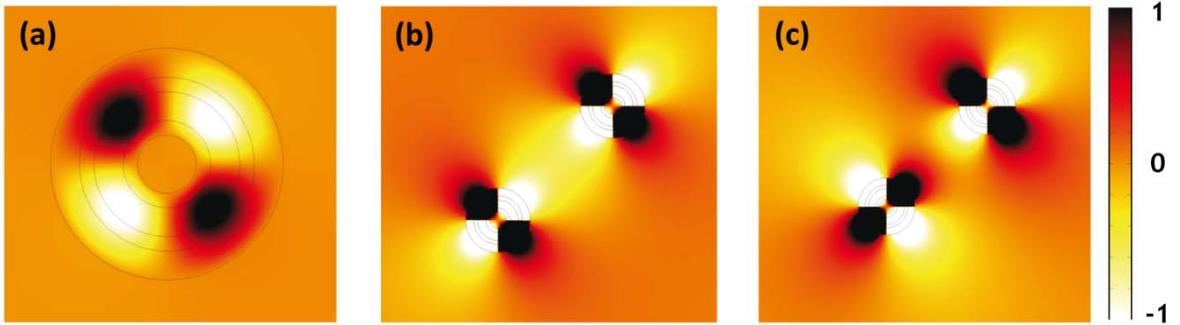

Fig. 2. Simulated electric field patterns in normalized units of (a) the quadrupolar resonance of a single RPC at $f_{q=2}$ = 179.33 MHz, and (b), strongly coupled even resonance pattern of a system formed by 2 resonant shells separated a distance (center to center) $D$ = 200 mm ≡ $5r_{ext}$. Resonance frequency for the even mode is 179.32 MHz, while it is 179.34 MHz for the odd mode in (c).

Figure 2(a) plots the selected quadrupolar resonance mode, at which the free space wavelength is close to $\lambda_0$ = 1.67 m. This selection is supported on the base that it provides a high $Q$-factor at the lowest possible frequency. By this means we have a tradeoff between the electrical size of the resonator and the intensity of the resonance. We have calculated the quality factor considering only radiation losses at $f_{q=2}$ as $Q$ = 4.42×10$^5$. Let us point out that the value of the total $Q$-factor is strongly dependent on the absorption or dissipative losses in the materials forming the shell. If typical dissipation losses for low loss materials are included (for example $tan\ \delta$ = 10$^{-4}$ for the dielectrics), the resulting $Q$-factor including radiation and absorption losses is $Q$ = 9779.64.

Although this value is two orders of magnitude lower than the previous one, it is for instance five times higher than that reported in [6] for similar absorption loss parameters ($Q$ = 1661). The discussion about the different sources of loss is in our opinion relevant, since absorption losses are inherently linked to the actual implementation scheme employed. For example, using resonant elements to implement the constitutive parameter profiles may add a significant amount of losses, so other schemes should be preferable.

Since the purpose of our study is to evaluate the possibilities of using metamaterial shells as a potential structures to perform wireless energy transfer no particular implementation scheme is here proposed so far, and no performance optimization targets are followed.

As the second step in the analysis we have studied the system of two coupled shells, see Figs. 2(b) and 2(c). In this case, two identical shells are located at moderate and variable distances ranging from $D/r_{ext}$ = 3 to $D/r_{ext}$ = 10. With the actual shell dimensions these distances vary from 12 cm to 40 cm. The combination formed by these two resonant elements creates a system with its own resonant modes and frequencies, which result from the original resonant modes of the individual element. A frequency split is characteristic of the interaction between both elements of the system, which in turn permits to evaluate the coupling rate or factor between them. It is evident that at shorter separations, the interaction between the resonant elements is higher and the frequency splitting is enhanced. Focusing on the quadrupolar resonance two modes are investigated, the even mode as in Fig. 2(b), and the odd mode as in Fig. 2(c). Both modes share the symmetry axis formed by the line equidistant from both centers of the RPCs. The purpose of the analysis of the combined system is to evaluate if a strong coupling regime is established between both resonant elements that can favor a wireless energy transfer. Two key parameters can be identified in order to evaluate such an interaction. On the one hand it is the coupling factor or rate $\kappa$ between the RPCs. This coupling factor relates the variations of the field amplitude in the first shell $a_1$ with the field amplitude in the second shell $a_2$ that have to satisfy the following conditions (assuming they both have identical resonant frequencies $\omega$ and damping factors $\Gamma$):

$$\frac{da_1}{dt} = -i(\omega - i\Gamma)a_1 + i\kappa a_2 \quad (5)$$

$$\frac{da_2}{dt} = -i(\omega - i\Gamma)a_2 + i\kappa a_1 \quad (6)$$

The coupling rate $\kappa$ can be evaluated in practice from the frequency separation between even and odd modes, since it is related to the intensity of the interaction. We have used the standard definition $\kappa = (\omega_o - \omega_e)/2$, where $\omega_o$ and $\omega_e$ are the angular frequencies of each mode. On the other hand, there is the rate at which energy is lost in the resonators, which can be quantified by means of the damping factor. In our case, we use the damping factor $\Gamma$ related to the $Q$ value and giving the width of the resonances. From the definition of the $Q$-factor, we have $Q = \omega/2\Gamma$. Since there are two resonant modes (even, odd), the averaged damping factor is $\Gamma = \sqrt{\Gamma_e \Gamma_o}$. In a practical situation, the coupling rate $\kappa$ should be sufficiently larger with respect to the damping factor $\Gamma$ in order to assure that energy is transferred at a rate higher than it is lost in the system.

A figure of merit (FOM) can be defined as the coupling to loss ratio $\kappa/\Gamma$, which has to reach values in principle much higher than 1. Our study consisted in analyzing the resonant system formed by two shells with split resonant frequencies. For each mode, even and odd, a coupling to loss ratio has been calculated from the respective frequencies and $Q$-factors. The results are summarized in Fig. 3, where panel (a) displays the resonant frequencies as a function of the distance between shells. It is observed that, as expected, $f_e$ and $f_o$ are quite different for the shorter distances and both tend to the single shell resonance frequency as distance increases. Let us recall that these are relative distances larger than the size of the shells, and at the same time they are smaller than the operation wavelength in free space.

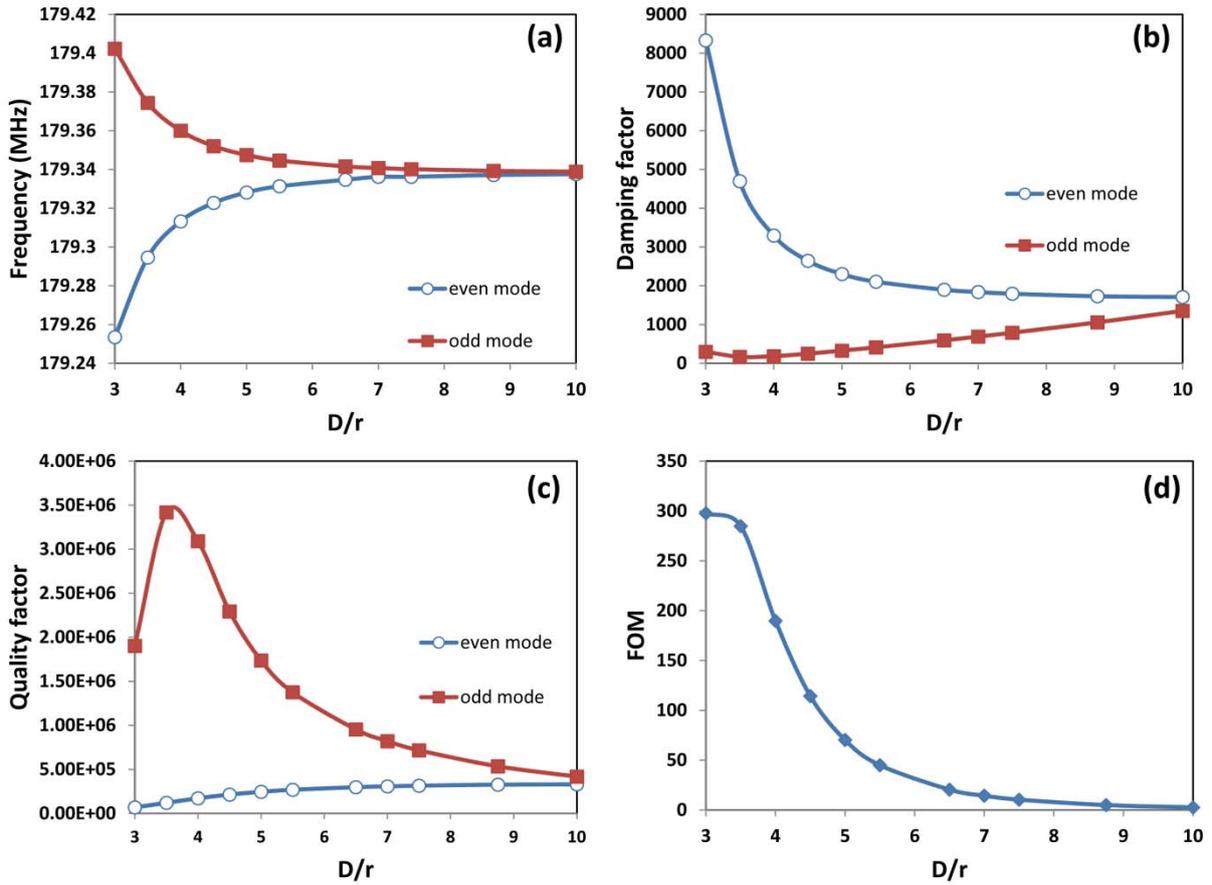

Fig. 3. Numerical results from an eigenvalue analysis of a pair of coupled resonators with varying separation distance $D/r_{ext}$ (ratio between the physical distance among their centers and the radius of the RPCs). (a) Resonant degenerated frequencies of the coupled RPCs system, (b) Damping factor for each resonant frequency including radiation losses (c) Calculated quality factors for both resonant modes. (d) Figure of merit (FOM) for the energy transfer calculated as the ratio between the coupling factor $\kappa$ and the averaged damping factor $\Gamma$.

Figures 3(b) and 3(c) report the quality $Q$ and damping $\Gamma$ factors (related only to radiation losses), respectively. The even mode presents more radiation losses, i.e. lower $Q$-factors. For the odd mode and since the damping factor presents a minimum value, a maximum value is obtained for the $Q$-factor. This maximum value corresponds to a separation close to four times the radius of the shells. This is an important difference between both resonant modes.

Finally, Fig. 3(d) displays the calculated coupling to loss ratio $\kappa/\Gamma$. Within a wide range of distances the FOM reaches values much higher than one. All this distance range where the FOM presents values higher than one is in principle usable in view of performing a wireless power transfer. Energy should be transferred at a rate higher than the rate at which it is lost in the system, which is a target in practice.

Additionally, we have compared the efficiency of the proposed resonator structures, based on metamaterials, with other solutions already explored. Thus, the analysis described in the manuscript has been also carried out for homogeneous isotropic disks with the same physical dimensions of the metamaterial shells. Specifically, the constitutive parameters are the result of averaging those of the RPC metamaterial shell: $\varepsilon_H$ = 30.06, $\mu_H$ = 33.23. Under these conditions, the individual resonator frequency is $f_{q=2}$ = 190.09 MHz and the associated quality factor is $Q_H$ = 1.42×10$^5$. The best coupling obtained is $FOM_H$ = 114.43 at $d$ = 12 cm.

The full results of FOM as a function of the separation distance are displayed in Fig. 4. They demonstrate that the coupling factor between the two anisotropic metamaterial shells is higher than that for the equivalent homogeneous dielectric and dielectric-magnetic disks. With the metamaterial resonators, higher quality factors can be obtained and thereby higher FOM (FOM$_{RPC}$ ≫ FOM$_H$). The proposed resonator structures based on metamaterials present much better performance for wireless energy transfer.

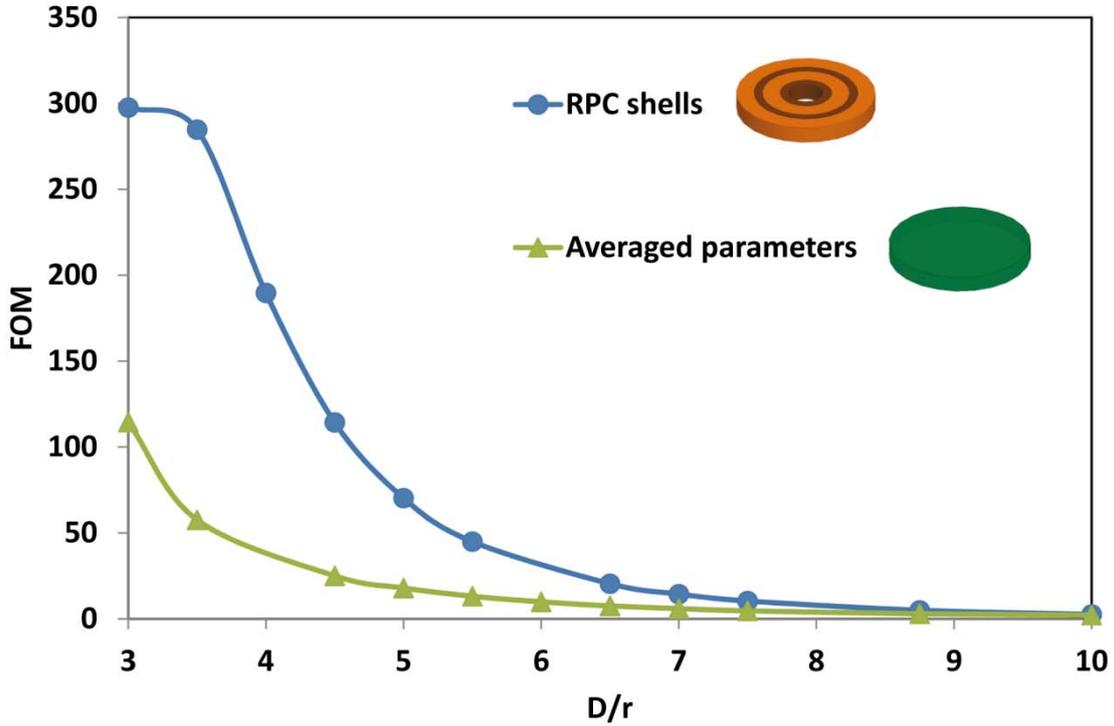

Fig. 4: Simulated coupling to loss ratio FOM = $\kappa/\Gamma$ for the metamaterial shells compared to homogeneous disks as a function of the normalized separation in each case.

A practical application of this type of system would require inevitably additional considerations in order to assess a magnitude quantifying the transfer efficiency. For this goal, we have analyzed the problem of placing two shells in proximity under conditions that could be encountered in practice. In particular, we have analyzed the possibility of using connecting elements respectively acting as a power source and a power drain. Therefore, 3D numerical simulations including feeding and probing coaxial connectors in both shells are performed. This permits to directly obtain the power transfer efficiency figure evaluating the performance of the system. This efficiency can be estimated directly from the transmission coefficient ($S_{21}$) relating the power from the transmitting port that is transferred to the receiving port. The system analyzed is composed of two identical devices including a coaxial standard connector on the shells inner cavities and each one acts respectively as power source and power drain.

A schematic of this configuration is given in Fig. 5. Importantly, the connectors themselves are part of the power transfer system and influence its performance. Also, the estimated efficiency by this means is a global value that includes the efficiency of the transfer between the connectors and the shells on top of the transfer efficiency between the two shells. Let us mention that the placement or geometrical design of these connecting elements is not optimized. Impedance adaptation between coaxial probe and the cavity material is the only consideration taken into account. Coaxial probes are not placed exactly at the center of the inner cavities. This is done in order to improve the excitation of the quadrupolar mode [13]. Since this mode has a null at the center of this cavity because of symmetry reasons, the center is not an optimum position for the source or load connectors. The optimal position to maximize matching between the connector and the shell has not been investigated and is therefore susceptible of improvement. In our simulations, the inner coaxial conductor is just displaced at $r = 0.1 r_{ext}$ from the center of each RPC. The presence of the connectors creates a slight perturbation in the electrical behavior of the RPCs, shifting in practice their resonant frequencies from the ones displayed in Fig. 2(a). However, the simulations show that this shift is very small (<0.01%) and it is almost homogeneous within the considered range of separation distances.

Figure 5 shows the evolution of the transfer efficiency as a function of the separation between the shells in normalized units. At each separation distance, the maximum transmission frequency is shifted (as it was already shown in Fig. 3(a)). The maximum transmission coefficient is obtained at each one of these peak frequencies: this maximum $S_{21}$ value is the one reported in Fig.4. It can be seen that efficiency is higher for the shorter distances and it slowly decays with increasing separations. Maximum transfer efficiency close to 83% is obtained for a separation of four times the device radius. This distance value is slightly above the one reported in Fig. 3(d), because the system has been altered due to the presence of the coaxial connectors and, as it was said previously, resonance frequencies have also been shifted. Efficiency remains high (> 35%) for separations up to ten times the radius of the shells. In this sense, it is interesting to note that this separation range corresponds, in terms of electrical distances to $D/\lambda = 0.07$ (for the shortest separation) to $D/\lambda = 0.22$ (for the largest separation). Actually, this means that in all cases we are at a short electrical distance, within the first quarter-wavelength away from the transmitting device. Hence, even reactive power can be used to couple energy from one device to the other.

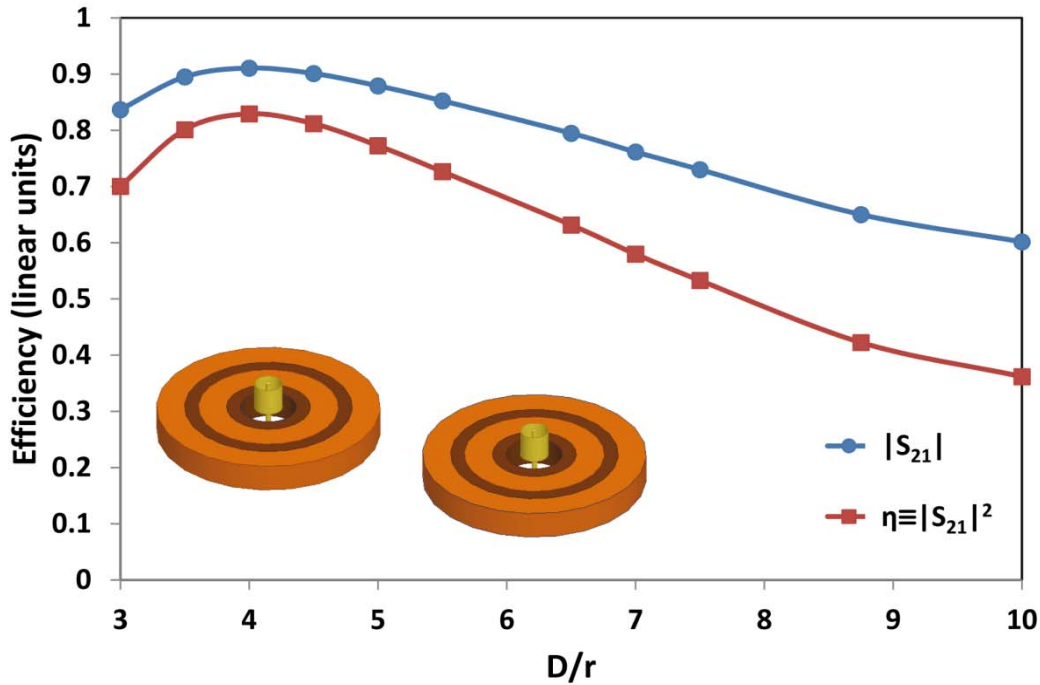

Fig. 5. Simulated power transfer efficiency $\eta$ and transmission coefficient $S_{21}$ as a function of separation distance. These relative figures are defined between the two port planes of the coaxial connectors. Efficiency $\eta$ includes the transfer rates from the connectors to the RPCs and the wireless transfer rate between the RPCs. Inset shows a schematic of the wireless energy transfer system including source and drain coaxial connectors to respectively inject and extract EM energy.

## 4. Conclusion

In conclusion, we have studied the behavior of Radial Photonic Crystals shells as useful structures to perform wireless power transfer. The proposed design is based on a high quality factor sub-wavelength resonator with appropriate constitutive parameters. The simulations indicate that this type of microstructure enhances the wireless transmission of power. Interaction of non-radiative fields in a strong coupling regime permits obtaining high power transfer efficiencies (up to 83%). We hope that our results motivate further practical developments on the basis of the proposal developed here.

**Acknowledgment**

This work was supported by the Spanish Ministry of Economy and Competitiveness under Grants TEC 2010-19751 and Consolider CSD2008-00066. The authors also acknowledge Daniel Torrent for useful discussions.